\documentclass{CIRED-v1}

\usepackage{subcaption}
\usepackage[utf8]{inputenc}
\usepackage{xurl}

\begin{document}

\title{DATA QUALITY CHALLENGES IN \\EXISTING DISTRIBUTION NETWORK DATASETS}

\author{Frederik Geth\ad{*1}, Marta Vanin\ad{2,3}, Dirk Van Hertem\ad{2,3}}

\address{
\add{1}{GridQube, Brisbane, Australia}
\add{2}{Department of Electrical Engineering, KU Leuven, Leuven, Belgium}
\add{3}{EnergyVille, Genk, Belgium}
\email{frederik.geth@gridqube.com}}


\begin{abstract}
\textit{Existing digital distribution network models, like those in the databases of network utilities, are known to contain erroneous or untrustworthy information. 
This can compromise the effectiveness of  physics-based engineering simulations and technologies, in particular those that are needed to deliver the energy transition.
The large-scale rollout of smart meters presents new opportunities for data-driven system identification in  distribution networks, enabling the improvement of existing data sets. 
Despite the increasing academic attention to  system identification for distribution networks, researchers often make troublesome assumptions on what data is available and/or trustworthy. 
In this paper, we highlight some differences between academic efforts and first-hand industrial experiences, in order to steer the former towards more applicable research solutions.}
\end{abstract}

\maketitle

\section*{INTRODUCTION}

The integration of renewables, batteries, heat pumps and electric vehicles in power distribution networks (DN) can be approached in a model-driven fashion. 
Mathematical optimization-based techniques like four-wire unbalanced optimal power flow can serve as basis for a variety of decision support tools in the context of operating and planning the DN, including but not limited to: state estimation (SE), battery dispatch optimization, dynamic operating envelopes for PV systems, etc. 
Around the world, trials are being run to demonstrate the value of such advanced network tools and there is a wealth of startups commercializing such technologies. 

Using electrical engineering models as the foundation for these tools offers advantages like explainability, granularity and ability to be validated. 
Being physics-based, these models are also non-biased. 
The key hurdle is the non-trivial input required: a valid network model including connectivity/topology and line/cable/transformer impedance information. 
Applying even the most advanced decision support tools to inaccurate network data inevitably leads to problematic or unreliable results. 
This hampers the adoption of new methods in the field, leading to subpar investment policies and slowing down the energy transition. 
Data capture and integration standards of the past are no longer sufficient to run in modern tools. 
Therefore, substantial data cleaning is required.

Recognizing that there are data quality issues, and identifying which specific ones, is itself not a trivial task.
SE engines are generally able to identify specific buses and sensors with significant mismatch between models and reality, and can be used to validate the correction/identification of DN data. 
Using system identification methods, one can analyse the accuracy of physical attributes such as impedance data and topology.

The high potential of DN data has been acknowledged by Energy Networks Australia (ENA) in the `Data opportunities for smarter networks' report~\cite{ENEAENA}.
The CSIRO-ENA network transformation roadmap \cite{entr} states that the full potential of distributed energy resources (DER) can only be realised in a future with multidirectional exchanges of both energy \emph{and information}. 
The Sandia report~\cite{osti_1855058} explores the use of measurement data to calibrate DN models for improved planning and grid integration of solar in the US. 
Note that differences in DN design patterns across the world affect the type of data quality issues and the methods to solve them. 
This paper reviews data quality issues that the authors observed in real-world network data from distribution utilities in Belgium and Australia: Fluvius and Ergon Energy Network and Energex, as well as the CSIRO LV feeder taxonomy~\cite{TaxonomyStudy}. Multiple other parts of the world likely present similar ones.
Different categories of issues can be established, that impact physics-driven decision support tools as a whole:
\begin{enumerate}
    \item \emph{modelling errors} or shortcuts due to applying the circuit laws under invalid assumptions, e.g., using Kron's reduction (KR) in networks where neutrals aren't pervasively grounded;
    \item \emph{network data errors}, e.g., problematic impedance values or topology information;
    \item (largely) inevitable \emph{measurement errors}: `noisy' or `bad' data due to sensor tolerances or malfunction; and
    \item \emph{measurement inadequacy} due to either A) \textit{semantic} mismatch, e.g., having averaged instead of instantaneous values, or rms magnitudes instead of base frequency magnitudes; B) \textit{granularity} mismatch, e.g., aggregated three-phase  measurements instead of per-phase; C) \emph{label} mismatch, e.g., wrong location or phase meta-data. 
\end{enumerate}
To improve the data quality long-term, we need to:
\begin{itemize}
    \item understand, apply and improve  best practices in network data set development;
    \item develop and use tools that automate data cleaning and maintenance tasks; and
    \item devise practical methods to validate data corrections using real-world observations.
\end{itemize}
Note that the higher quality the obtained network model is, the more use cases are possible. For example, explicit grounding models allow fault analysis in addition to power flow analysis. Similarly, modelling wire types and geometry for lines/cables allows harmonic analysis to be performed by solving Carson's equations for different frequencies.

\vspace{2mm}
\noindent{\underline{\textbf{Scope and contributions}}}
\vspace{1.5mm}
 
Many scientific papers rely on data and assumptions affected by the four sources of errors for decision support tools  presented in the previous section. 
Urquhart et al.~\cite{Urquhart2013} review assumptions and approximations that are typically applied in LV network research. 
Most of these are purely modelling shortcuts (item {1}). 
 Blakely et al.~\cite{8981211} discuss considerations that utilities need to
make when implementing data collection policies (item {4}), and propose some techniques to address them. 
Conversely, our work focuses on  \emph{network data errors} (item 2). 

The goal is to improve awareness of real-world data issues in the academic community, to foster the development of physics-driven frameworks that identify such issues and incrementally improve untrustworthy data sets.
We categorise the issues, present some of the historical reasons behind them, and illustrate examples from real-world experiences. 

\section*{DATA PRACTICE IN DISTRIBUTION}
Advanced Distribution Management Systems (ADMS) are often envisioned to become the key platform for network operations. 
Dubey et al. note that ADMS applications will benefit from  data-centric components to quantify and improve the quality and consistency of the network data \cite{dubey2020}.
Today, however, many of the data sources have limited integration with respect to electrical engineering use cases, with a lack of consistency checks and lack of continual synchronization.
Commonly, the data architecture may look as:
\begin{itemize}
    \item The geographic information system (GIS) tracks the locations of network assets. Lines are tracked as paths, and switches/breakers and transformers are commonly represented as nodes. The \emph{states} of switches/breakers are not tracked as part of this data set. 
    \item SCADA sensor data flows into time series databases. SCADA (sensors) may be set up to only communicate measurements when certain thresholds for change are exceeded, thereby limiting resolution.
    \item (A)DMS is used to track the on-line configuration of substations, so crews can be informed about live circuits risks, but it is less frequently used for power flow analysis. 
    \item Asset databases for instance track nameplates, spec sheets, procurement and field installation details. 
    \item A DER register tracks the installed capacities of solar, batteries and more. This is generally not integrated with smart meter (SM) databases, so checking for inconsistency is hard.
    Furthermore, not all jurisdictions require customers to inform their utility company about the installation of certain inverter-based resources.
    \item SM data, both related to billing information and engineering data, is stored in separate time series databases. In Australia, third parties frequently often own the data - not the utility. 
    \item Electrical engineering tools like PowerFactory, PSS/Sincal or even in-house ones, are used for design and planning purposes. Their input models are often based on data exports from GIS, after which some data gaps are filled based on engineering standards or experience. These pieces of tacit knowledge are not pushed to other information systems. 
    \item Mappings of line construction codes to impedance values are sometimes maintained in a separate database. 
\end{itemize}
A comprehensive example of a real-world data architecture is given in EPRI technical brief \cite{epri_gmdm} (p.6).
The lack of integrated databases can cause data management difficulties, and creates data quality issues when combining different sources for engineering models. 
Some of these errors can be spotted with SM data, e.g., failure to report a PV installation from the register to the engineering model can be detected by presence of power injection. 
The errors described in this section add to those from wrong modelling error choices, like KR, and those induced by the individual unknowns/errors in the separate databases, e.g., impedance and phase connectivity information.

Practitioners need to be aware of the `best practices' in engineering modelling. 
Utilities may have different teams responsible for modelling subtransmission, MV and LV networks. 
Phase unbalance is easily neglected when taking high-voltage network modelling approaches and applying them to MV networks. Similarly, neglected neutral voltage rise and sequence impedance parameterizations may occur due to the adoption of MV modelling approaches in LV contexts. These are avoidable approximations.

It is interesting to contrast academic perspectives with industrial ones. 
A blog post\footnote{\url{https://energycentral.com/c/ua/asset-data-quality-challenges-and-opportunities-transmission-and-distribution}} discusses a number of organizational obstacles to the adoption of better digital network models:
\begin{itemize}
    \item lack of accountability and responsibility in developing and maintaining data sets;
    \item excessive data flows from smart devices and sensors;
    \item field teams with high degrees of tacit knowledge that do not trust the data in the IT systems;
    \item lack of understanding with administrators on how the DN is constructed and operated; and
    \item product issues in managing and maintaining data.
\end{itemize}

\section*{ISSUES IN REAL-WORLD DATA}

(Quasi-)real-time detection of topology changes is part of transmission system SE. 
However, SM measurement \textit{time-series} can be used \cite{Subasic2022} to calibrate network topology for real-time use, improving the existing datasets at the same time. 
The Sandia report~\cite{osti_1855058} develops methodologies for phase identification, meter-to-transformer mapping, identification of voltage regulators, PV systems and their parameters, and many more. 

\begin{figure}[tbh]
   \includegraphics[width=2.2in, height=1.4in]{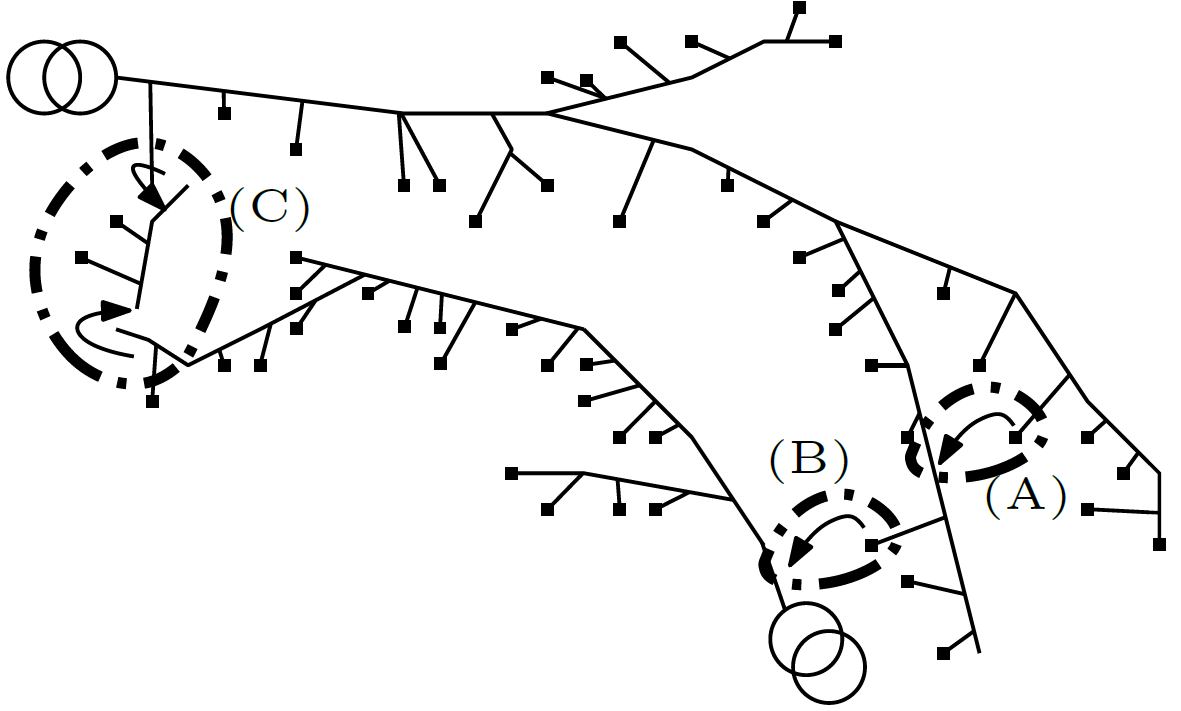}
   \centering
   \caption{Examples of possible user-to-cable errors in LVDN data. The user in (A) is
actually connected to the wrong branch of the same feeder, that in (B) is
connected to the wrong feeder altogether. Finally, the switch status (C) is is
wrong, and the two users between the switches are assigned to a wrong feeder.}
   \label{fig:feeder_errors}
\end{figure}

Table~\ref{tab_data_issues_martav2} lists a number of DN data issues, together with a summary of their impact on decision support methods. We find that the following are broadly under-addressed:
\begin{enumerate}
    \item Meter-phase alignment for three-phase residential consumers, as well as transformer/breaker monitors. 
    \item Some topology errors: incorrect meter-to-transformer assignments (see Fig.~\ref{fig:feeder_errors} errors (B)-(C)), missing information on single-phase branches off the main feeder, missing information on neutral grounding points.
    \item Transformer and regulator models: wrong nominal voltage rating (e.g., 433 vs 415 vs 400 V phase-to-phase for European style 3-phase grids) frequently caused by semantic confusion related to `voltage levels', and more broadly missing parameters of winding configuration, vector group and impedance.
    \item Missing detail on DER smarts, e.g. PV and battery control.
\end{enumerate}
Furthermore, combinations of multiple error sources in the models are also under-addressed, while in our industrial experience, we observed that these are likely and can be critical. For instance, phase identification methods perform noticeably worse if topology errors are present in addition to unknown/wrong phase assignment. 

In the upcoming sections we elaborate on the most common DN model errors and point out some relevant literature. 

\begin{table*}[tb]
\renewcommand{\baselinestretch}{0.7}
\centering
\caption{Data issues, an asterisk indicates particularly under-addressed ones.} \label{tab_data_issues_martav2}
{\footnotesize
\begin{tabular}{l p{0.3\textwidth}  p{0.6\textwidth} } 
\hline
& Issue & Implications and risks \\ \hline
1 & Missing phase labels (loads, transformers, etc.) & Inaccurate per-phase power flows, voltage unbalance inexact. \\
2 & Switch states \& user-branch connections & See issue 1 + impossible to determine how many consumers are connected to a transformer (aggregate values at feeder head inaccurate). \\
3* & Meter-to-transformer assignment & See issue 2. \\
4* & Excessive number of buses & Errors if reductions are performed, increased computational time. \\
5 & Wrong transformer tap & Inaccurate estimates of: end-consumer voltages, hosting capacity, voltage-based curtailment.\\
6 & Missing tap semantics & Taps are usually given integer values, the nominal value is generally not 0, the tap percentage can be unknown, see issue 5.\\
7* & Unknown vector group usage (ansi vs euro) & Sign error in angle off-set between primary and secondary for Dy transformers. \\
8* & Unknown winding configuration & Inaccurate voltage values, problematic for harmonic studies. \\
9* & Mislabeled transformer primary/secondary nominal voltage & See issue 5 + this also implies wrong transformer impedance (as they are typically specified in per unit w.r.t. the transformer power rating and primary voltage).\\
10* & Wrong transformer rating & Inaccurate classification of congestion; also implies inaccurate transformer impedance (as they are typically specified in per unit w.r.t. the transformer power rating and primary voltage).\\
11* & Only Kron-reduced impedance matrices available & Assumes neutral voltage rise is marginal under normal conditions, which is likely problematic in sparsely grounded networks. Not compatible with short circuit analysis. \\
12* & Only sequence impedances available & Equivalent to assuming transposition + multi-grounding of neutral in 4-wire networks. \\
13* & Meter-phase-alignment for three-phase users & Inaccurate SE and other measurement-based computations. \\
14* & Missing information on neutral grounding & Inaccurate voltage and current estimates, impact depends on grounding philosophy. \\
15 & Load model (constant power, ZIP) unknown & Analyses may be inaccurate in terms of voltages, unbalance levels, neutral current.\\
16 & Missing load (locations) & SE may increase load at the known locations, leading to inaccurate congestion identification. \\
17 & Measurements rms vs fundamental-only & When using a fundamental-frequency-only (O)PF or SE, contributions of the higher frequencies to the RMS values may lead to inaccuracy. \\
18* & Regulators modelled as transformers & Impedance values differ between transformers and regulators, inaccurate  voltages/currents. \\
19 & Missing capacitor banks (specifications) & Reactive power flows / power factor for loads/generators look statistically unlikely.  \\
20 & Approximate cable/line impedance models & Inaccurate estimation of currents (particularly neutral current), and voltage drops.\\
21* & Unknown (PV) inverter settings & Constant power factor and volt-var/watt lead to very different patterns of overvoltage. \\
22* & Unknown home battery dispatch strategies & Complementarity between PV, batteries \& load overestimated, inaccurate  voltages/currents.
\\
 \hline
\end{tabular}
}
\end{table*}

\vspace{2mm}
\noindent{\underline{\textbf{Phase labels and meter-to-transformer assignments}}}\label{sec:phase_labels}
\vspace{1.5mm}

Phase labels are \emph{essentially always unknown} and phase identification has received considerable research attention~\cite{Hoogsteyn2022, Bariya}. 
In our industrial experience, to obtain usable models for power flow calculations, utilities assign arbitrary values to the phase connections, e.g., assuming that users are divided equally among the phases. 
In single-phase DNs, like North-American LV ones, the equivalent exercise  is the meter-to-(single-phase) transformer assignment. 
In three-phase DNs, wrong user-to-transformer mappings affect phase and topology identification.
\noindent{\underline{\textbf{Network layout}}}\label{sec:layout}
\vspace{1.5mm}

Network layouts are usually known to a significant extent: line paths are derived by interpolating GIS objects (transformers, buildings, etc.). 
However, errors like (A) and (B) in Fig.~\ref{fig:feeder_errors} are common, especially where the objects are physically close. 

Improvements may be possible on the criteria that assign users to feeders and branches. 
For instance, a criterion used in Belgium is to assign an object to the network branch which is closest to the object's geometrical baricenter. A better rule-of-thumb would be that of using the distance from the front door. 
Nevertheless, in LV networks we observed that user-cable connections are occasionally `illogical' from an electric installation perspective, e.g., users are not connected to the closest feeder cable due to historical circumstances. 

Inconsistencies in the DN layout fall under the category of `topology errors'. 
Errors like Fig.~\ref{fig:feeder_errors} (B) are typically easier to detect than errors like (A) by statistical analysis: load patterns, unbalance levels and tap settings cause larger voltage profile differences across different transformers than they do across different branches of the same transformer. A situation `between' (A) and (B), relatively common in Flanders, is to have a user connected to a different feeder from the same transformer.
Wrong switch states, like Fig.~\ref{fig:feeder_errors} (C), imply 
meter-to-transformer 
or meter-to-branch mismatches, and 
are caused by a lack of tracking of 
the DN layout after maintenance and operation actions. 

\begin{figure}[t]
\includegraphics[width=0.9\columnwidth]{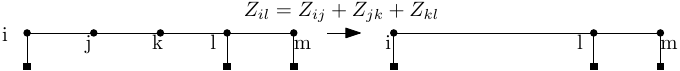}
   \centering
   \caption{Elimination of superfluous buses.}
   \label{fig:bus_reduction}
\end{figure}

Finally, GIS-derived models present a high number of electrically superfluous nodes. 
A well-known example is the IEEE European LV Test Feeder, which has 906 buses originally.
The buses can be reduced to less than 120 without approximation, by adding up the lengths of multiple segments of the same type as in Fig.~\ref{fig:bus_reduction}. 
Such reductions can be applied to any network data set, 
and results in streamlined data and improved solve times for physics-based optimization and simulation engines.

\vspace{2mm}
\noindent{\underline{\textbf{Transformer parameters}}}
\vspace{1.5mm}

The  tap setting of MV/LV transformers is frequently unknown. 
Tap positions can change on a seasonal basis or when voltage problems are reported, so they are often off-nominal. 
Tap settings and transformer parameters are crucial when  MV+LV models are set up.
Yusuf et al. \cite{yusuf2022} explore methods to infer operational taps of transformers and regulators.

Transformer impedances are generally specified in the transformer's own per-unit impedance base. Therefore, impedance values are inaccurate when any of the following are inaccurate: primary voltage, transformer power rating, or the impedance itself. Electrical lab tests are frequently used to identify missing electrical parameters. 

\vspace{2mm}
\noindent{\underline{\textbf{Cable/line properties and grounding}}}
\vspace{1.5mm}

Depending on the local practice, measurement devices on the secondary side of a substation may or may not be present. 
In our experience, combining transformer and digital meter measurements, e.g., for SE, can be nontrivial.
For instance, a neutral-voltage shift could be observed on the SM reading of a Belgian consumer, which was not observable at the transformer. 
Amongst the main factors that could cause this are 1) the fact that the transformer may be (nearly) perfectly grounded, while (metered) users are generally not in many places in Europe, 2) the lack of available grounding information. 
Under a sparse grounding philosophy, e.g., in Belgium, KR of the neutral is inappropriate, so four-wire models are crucial to capture neutral voltage shifts. 
In places with a pervasive neutral grounding philosophy, e.g., Australia, KR can be used, and design standards on grounding can be used to infer likely grounding locations and impedances. Note that applying KR to every segment of bus-rich four-wire networks (Fig.~\ref{fig:bus_reduction} left) adds modelling errors: superfluous buses are not grounded. 

Assuming a four-wire model is used, and neutral grounding is known and modelled, wrong impedance representations may still  cause errors. 
Utilities might not know what line types are placed in certain parts of the grid. 
Even when the type is known, chances are high that only the positive sequence and -- occasionally but not always -- zero sequence impedances are known. 
This is not enough information to recover an untransposed impedance model. If nominal cable impedances are known accurately, cable lengths are likely still approximate, as GIS segments do not exactly match cable paths. 
However, performing line length identification is easier than estimating the whole impedance model~\cite{Vanin2020}. 
Finally, shunt impedances are usually neglected, although in LV networks this appears acceptable~\cite{Urquhart2013}. 

Nameplates and spec sheets provide crucial parameters but generally not all. 
For instance, the impedance of an overhead line is a function of its geometry due to induction, not just of the wires used. 

Three solutions have been proposed to derive accurate, untransposed impedance values: 1) solving (modified) Carson's equations, 2) data- and physics-driven reconstruction from SM measurements~\cite{Vanin2020} and 3) finite-element methods to solve Maxwell's laws across sections of line/cable~\cite{urquhart2015series}. 

\section*{VISION AND FUTURE WORK}

We believe essentially all of the data issues can be overcome, and that they do not need to be a major hurdle in the deployment of better physics-driven support tools in the real world.
Nevertheless, bringing the deployment costs down further through novel methods is a major research opportunity.
\begin{figure*}[!h]
\includegraphics[width=0.80\textwidth]{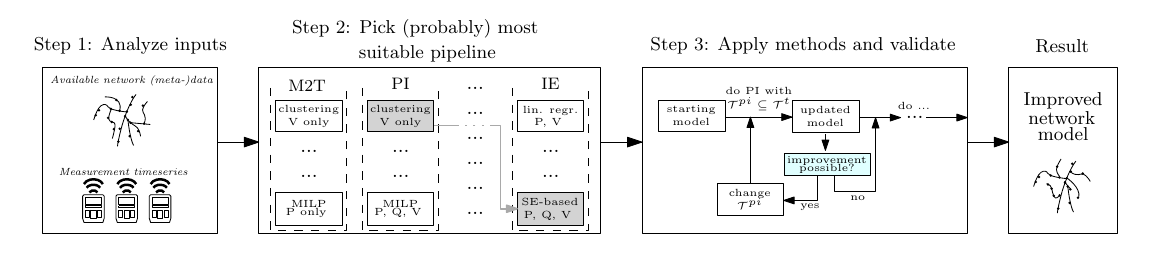}
   \centering
   \caption{Sequential, validated and automated data cleaning framework. (MILP: mixed-integer linear programming.)}
   \label{fig:pipeline}
\end{figure*}
Therefore, Fig.~\ref{fig:pipeline} sketches a system identification/network data cleaning framework to improve existing DN models that represents the authors' vision. 
Different calibration tasks are sequentially applied: meter-to-transformer (M2T) assignment, phase identification (PI), etc., forming the pipelines (Step 2). 
Given the wide variety of DN features (unbalance level, number of users, etc.), utilities would benefit from toolboxes that have different methods for each task. 
A robust framework makes an informed guess on which methods are preferred in a given context (analyzing inputs, Step 1). 
Features that affect the accuracy of a method need to be investigated: e.g., voltage clustering might not work well for PI in rather balanced DNs, whereas methods that can exploit both power and voltage measurements may be accurate. 
Similarly, the NYSERDA report~\cite{nyserda} highlights that in the identification and calibration of SE input errors, the error type is assumed to be known, and
 bad data 
have been removed a priori. 
The impact of bad data on system identification processes is underaddressed. 
Furthermore, data from during outages can be exploited to identify topology and phase inaccuracies. 

Finally, real-world-compatible validation strategies are needed 1) to make sure that the new model improves over the old one, and 2) to evaluate the performance of different pipelines, picking the one with the best result, and/or choosing the optimal training set to use (e.g., $\mathcal{T}^{pi}$ in Step 3 of Fig.~\ref{fig:pipeline}). 
Present system identification literature benchmarks the accuracy of the proposed methods to the `ground truth' of the synthetic networks used in the paper. 
This is not possible in real life and is a fundamental issue ignored by most papers, possibly due to lack of access to utility data. 

We believe that, to be effective, such frameworks must combine data science and physics-based methods, and  identify  opportunities in system identification research:
\begin{itemize}
    \item tackling the under-addressed network data error sources;
    \item handling multiple error sources in an integrated fashion;
    \item understanding and mitigating the impact of unfavourable measurement conditions on system identification methods;
    \item integrating real-world validation.
\end{itemize}

Furthermore, DNs are not static: once the present system is identified, models need to be kept up-to-date, consistent and interpretable. 
System changes must be identifiable with limited measurement requirements.
To all ends, researchers would benefit from the access to real-life utility data.
Recommendations for improved data and modelling practices include:
\begin{itemize}
    \item well-defined  semantics for the data models;
    \item consistency checking of data sets, including tagging whether network data has been Kron-reduced or not;
    \item user-friendly data debugging solutions. Power flow solvers throwing a `singularity error' is  unhelpful, and doesn't help the discovery and learning process.
\end{itemize}

\section*{ACKNOWLEDGEMENT}
 The authors thank Terese Milford and Wilson Bow at Ergon Energy Network and Energex, and Andy Gouwy and Bruno Macharis at Fluvius for their support and constructive feedback. 

\section*{REFERENCES}

\bibliographystyle{IEEEtran}
\footnotesize
\bibliography{library}

\end{document}